\documentclass[aps,prb,superscriptaddress,twocolumn,floatfix]{revtex4}     
\usepackage{epsfig,dcolumn,amsmath,latexsym}
\begin{document}

\newcommand{\IV}{$I$-$V$}

\title{Density functional method for nonequilibrium electron transport}

\author{Mads Brandbyge}
\email{mbr@mic.dtu.dk}
\affiliation{Mikroelektronik Centret (MIC), 
Technical University of Denmark, Bldg. 345E,
DK-2800 Lyngby, Denmark}

\author{Jos\'e-Luis Mozos}
\affiliation{Institut de Ci\`encia  de Materials de Barcelona, 
CSIC, Campus de la U.A.B., 08193 Bellaterra, Spain}

\author{Pablo Ordej\'on}
\affiliation{Institut de Ci\`encia  de Materials de Barcelona, 
CSIC, Campus de la U.A.B., 08193 Bellaterra, Spain}

\author{Jeremy Taylor}
\affiliation{Mikroelektronik Centret (MIC), 
Technical University of Denmark, Bldg. 345E,
DK-2800 Lyngby, Denmark}

\author{Kurt Stokbro}
\affiliation{Mikroelektronik Centret (MIC), 
Technical University of Denmark, Bldg. 345E,
DK-2800 Lyngby, Denmark}

\date{\today}

\begin{abstract}
  We describe an {\it ab initio} method for calculating the electronic
  structure, electronic transport, and forces acting on the atoms, for
  atomic scale systems connected to semi-infinite electrodes and with
  an applied voltage bias. Our method is based on the density
  functional theory (DFT) as implemented in the well tested {\sc
    Siesta} approach (which uses non-local norm-conserving
  pseudopotentials to describe the effect of the core electrons, and
  linear combination of finite-range numerical atomic orbitals to
  describe the valence states). We fully deal with the atomistic
  structure of the whole system, treating both the contact and the
  electrodes on the same footing.  The effect of the finite bias
  (including selfconsistency and the solution of the electrostatic
  problem) is taken into account using nonequilibrium Green's
  functions. We relate the nonequilibrium Green's function
  expressions to the more transparent scheme involving the scattering
  states.  As an illustration, the method is applied to three systems
  where we are able to compare our results to earlier {\it ab initio}
  DFT calculations or experiments, and we point out differences
  between this method and existing schemes. The systems considered
  are: (1) single atom carbon wires connected to aluminum electrodes
  with extended or finite cross section, (2) single atom gold wires,
  and finally (3) large carbon nanotube systems with point defects.
\end{abstract}


\maketitle
\section{Introduction}
\label{sec:intro}

Electronic structure calculations are today an important tool for
investigating the physics and chemistry of new molecules and
materials.\cite{Fu95} An important factor for the success of these
techniques is the development of first principles methods that make
reliable modeling of a wide range of systems possible without
introducing system dependent parameters.  Most methods are, however,
limited in two aspects: (1) the geometry is restricted to either
finite or periodic systems, and (2) the electronic system must be in
equilibrium.  In order to address theoretically the situation where an
atomic/molecular-scale system (contact) is connected to bulk
electrodes of requires a method capable of treating an infinite and
non-periodic system. In the case where a finite voltage bias applied
to the electrodes drives a current through the contact, the electronic
subsystem is not in thermal equilibrium and the model must be able to
describe this nonequilibrium situation.
The aim of the present work is to develop a new first principles
nonequilibrium electronic structure method for modeling a
nanostructure coupled to external electrodes with different
electrochemical potentials (we will interchange the terms {\em
  electrochemical potential} and {\em Fermi level} throughout the
paper).  Besides, we wish to treat the whole system (contact and
electrodes) on the same footing, describing the electronic structure
of both at the same level.  

Our method is based on the Density Functional Theory
(DFT).\cite{HoKo64,KoSh65,KoBePa96,PaYa89} In principle the exact
electronic density and total energy can be obtained within the DFT if
the exact exchange-correlation (XC) functional was available. This is
not the case and the XC functional has to be substituted by an
approximate functional.  The most simple form is the Local Density
Approximation (LDA), but recently a number of other more complicated
functionals have been proposed, which have been shown to generally
improve the description of systems in equilibrium.\cite{KuPeBl99}
There is no rigorous theory of the validity range of these functionals
and in practice it is determined by testing the functional for a wide
range of systems where the theoretical results can be compared with
reliable experimental data or with other more precise calculations.

Here we will take this pragmatic approach one step further: We will
use not only the total electron density, but the Kohn-Sham wave
functions as {\it bona fide} single-particle wave functions when
calculating the electronic current. Thus we assume that the commonly used XC
functionals are able to describe the electrons in non-equilibrium
situations where a current flow is present, as in the systems we wish
to study.\cite{cdft} This mean-field-like, one-electron approach is not able to
describe pronounced many-body effects which may appear in in some
cases during the transport process.  Inelastic scattering {\it e.g.}
by phonons\cite{NeShFi01} will not be considered, either.

Except for the approximations inherent in the DFT, the XC functional, and 
the use of the Kohn-Sham wave functions to obtain a current, 
all other approximations in the method are controllable, in the sense that they
can be systematically improved to check for convergence towards the
exact result (within the given XC functional). Examples of this are
the size and extent of the basis set (which can be increased to
completeness), the numerical integration cutoffs (which can be
improved to convergence), or the size of the electrode buffer regions
included in the selfconsistent calculation (see below).  
This mean-field-like, one-electron approach is not able to
describe pronounced many-body effects which may appear in in some
cases during the transport process.  Inelastic scattering {\it e.g.}
by phonons\cite{NeShFi01} will not be considered, either.

Previous calculations for open systems have in most cases been based
on semi-empirical
approaches.\cite{PeMaFl90,TiDa94,ChSaMu98,PaHeMiLeMaFl98,EmKi99,BrKoTs99,MeTaGuWaRo00,SaJo92,HaReHuSi00,MuRoRa00,BiBuKe95,NeFi99,Na99,NaBe99}
The first non-equilibrium calculations with a full self-consistent DFT
description of the entire system have employed the jellium
approximation in the electrodes.\cite{La95,HiTs95} Other approaches
have used an equilibrium first principles Hamiltonian for the
nanostructure and described the electrodes by including semi-empirical
selfenergies on the outermost
atoms.\cite{YaRoGoMuRa99,DeSe01,PaPeLoVe01} Lately, there have been
several approaches which treat the entire system on the same
footing, at the atomic level\cite{WaMoTa97,ChIh99,CoCeSa99} but so far 
only one of the approaches have been applied to the nonequilibrium 
situation where the external leads have different electrochemical
potentials.\cite{TaGuWa01a,TaGuWa01b}

The starting point for our implementation is the {\sc Siesta} electronic
structure approach.\cite{SaOrArSo99} In this method the effect of the
core electrons is described by soft norm-conserving
pseudopotentials\cite{TrMa91} and the electronic structure of the
valence electrons is expanded in a basis set of numerical atomic orbitals
with finite range.\cite{SaNi89,ArSaOrGaSo99} The quality of the
basis set can be improved at will by using
multiple-$\zeta$ orbitals, polarization functions, etc.,\cite{ArSaOrGaSo99}
allowing to achieve convergence of the results to the desired
level of accuracy.
{\sc Siesta} has been tested in a wide variety of systems,
with excellent results.\cite{Or00,Oretal01}
The great advantage of using orbitals with finite range
(besides the numerical efficiency\cite{SaOrArSo99}), is that the
Hamiltonian interactions are strictly zero beyond some 
distance, which allows to partition the system unambiguously,
and define regions where we will do different parts of the calculation
as we describe in the Sections~\ref{sec:sysset}-\ref{sec:veff}.
Besides, the Hamiltonian takes the same form as in empirical tight-binding
calculations, and therefore the techniques developed in this
context can be straightforwardly applied.

We have extended the {\sc Siesta} computational package to
nonequilibrium systems by calculating the density matrix with a
nonequilibrium Greens functions
technique.\cite{HaJa96,Datta,BrKoTs99,TaGuWa01a} We have named this nonequilibrium
electronic structure code {\sc TranSiesta}.  Preliminary
results obtained with {\sc TranSiesta} were presented in
Ref.~\onlinecite{BrStTaMoOr01}.  Here we give a detailed account of
the technical implementation and present results for the transport
properties of different atomic scale systems. One of the authors (JT)
has been involved in the independent development of a package,
McDCAL,\cite{TaGuWa01b} which is based on similar principles, but with
some differences in implementation. We compare results obtained with
the two packages for a carbon wire connected to aluminum electrodes
and show that they yield similar results. We present new results for
atomic gold wire systems which are one of the most studied atomic
scale conductors, and finally we present results for transport in
nanotubes with defects.

The organization of the paper is the following. In the first part of
the paper we describe how we divide our system into the contact and
electrode parts and how we obtain the density matrix for the
nonequilibrium situation using Green's functions. Here we also discuss
the relation between the scattering state approach and the
nonequilibrium Green's function expression for the density matrix.
Then we describe how this is implemented in the numerical procedures
and how we solve the Poisson equation in the case of finite bias. In
the second part of the paper we turn to the applications where our aim
is to illustrate the method and show some of its capabilities rather than
presenting detailed analysis of our findings. We compare our results
with other {\it ab initio} calculations or experiments for (1) carbon
wires connected to aluminum electrodes, (2) gold wires connected to
gold electrodes, and finally (3) infinite carbon nanotubes containing
defects.

\section{System Setup}
\label{sec:sysset}

\begin{figure}[thb]
  \epsfig{figure=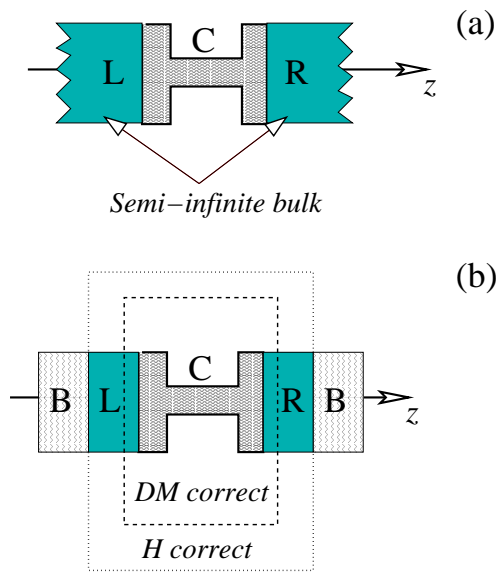, width=0.8\columnwidth, angle=0}
\caption{(a) We model the
  {\em Contact} ($C$) region coupled to two semi-infinite {\em Left}
  ($L$) and {\em Right} ($R$) electrodes.  The direction of transport
  is denoted by $z$. (b) We only describe a finite section of the
  infinite system: Inside the $L$ and $R$ parts the Hamiltonian matrix
  elements have bulk electrode values. The external (\em Buffer) region, $B$, is
  not directly relevant for the calculation.}
\label{fig:setup1D}
\end{figure}

We will consider the situation sketched in Fig.~\ref{fig:setup1D}a.
Two semi-infinite electrodes, left and right, are coupled via a contact
region. All matrix elements of the Hamiltonian or overlap integrals 
between orbitals on atoms situated in different electrodes are zero so the
coupling between the left and right electrodes takes place {\em via}
the contact region only.

The region of interest is thus separated into three parts, {\em Left}
($L$), {\em Contact} ($C$) and {\em Right} ($R$).  The atoms in $L$
($R$) are assumed to be the parts of the left (right) semi-infinite
bulk electrodes with which the atoms in region $C$ interact. The
Hamiltonian is assumed to be converged to the bulk values in region
$L$ and $R$ along with the density matrix.  Thus the Hamiltonian,
density and overlap matrices only differ from bulk values in the
$C$, $C-L$ and $C-R$ parts. We can test this assumption by including
a larger fraction of the electrodes in $C$ (so the $L$ and $R$ regions are
positioned further away from the surfaces in Fig.~\ref{fig:setup1D}).

In order to obtain the transport properties of the 
system, we only need to describe the finite $L-C-R$ part of the
infinite system as illustrated in Fig.~\ref{fig:setup1D}b.  The density
matrix which describes the distribution of electrons can be obtained
from a series of Green's function matrices of the infinite system as we will
discuss in detail in Section~\ref{sec:nedm}. In principle the
Green's function matrix involves the inversion of an infinite matrix
corresponding to the infinite system with all parts of the electrodes
included. We are, however, only interested in the finite $L-C-R$
part of the density matrix and thus of the Green's function matrix.
We can obtain this part by inverting the {\em finite} matrix,
\begin{equation}
\left(
\begin{tabular}{c|c|c}
${\bf H}_L + {\bf \Sigma}_L$ & ${\bf V}_L$ & $0$ \\ \hline
${\bf V}_L^\dagger$ & ${\bf H}_C$ & ${\bf V}_R$  \\ \hline 
$0$ & ${\bf V}_R^\dagger$ & ${\bf H}_R + {\bf \Sigma}_R$ \\
\end{tabular}
\right),
\label{eq:Hamil}
\end{equation}
where ${\bf H}_L$, ${\bf H}_R$ and ${\bf H}_C$ are the Hamiltonian matrices
in the $L$, $R$ and $C$ regions, respectively, and ${\bf V}_L$ (${\bf V}_R$)
is the interaction between the $L$ ($R$) and $C$ regions.
The coupling of $L$ and $R$ to the remaining part of the
semi-infinite electrodes is fully taken into account by the self-energies,
${\bf \Sigma}_L$ and ${\bf \Sigma}_R$.  We note that to determine ${\bf V}_L$,
${\bf V}_R$ and ${\bf H}_C$, we do not need to know the correct density
matrix outside the $L-C-R$ region, as long as this does not influence
the electrostatic potential inside the region. This is the case for
metallic electrodes, if the $L-C-R$ region
is defined sufficiently large so that all the screening takes
place inside of it.

The upper and lower part of the Hamiltonian (${\bf H}_{L(R)} + {\bf
  \Sigma}_{L(R)}$) are determined from two separate calculations for
the bulk systems corresponding to the bulk of the left and right
electrode systems.  These systems have periodic boundary conditions in
the $z$ directions, and are solved using Bloch's theorem. From these
calculations we also determine the self-energies by cutting the
electrode systems into two semi-infinite pieces using either the ideal
construction\cite{WiFeLa82} or the efficient recursion
method.\cite{LoLoRu84}

The remaining parts of the Hamiltonian, ${\bf V}_L$, ${\bf V}_R$ and
${\bf H}_C$, depend on the non-equilibrium electron density and
are determined through a selfconsistent procedure.  In Section~\ref{sec:nedm}
we will describe how the non-equilibrium density matrix can be
calculated given these parts of the Hamiltonian, while in
Section~\ref{sec:veff} we show how the effective potential and
thereby the Hamiltonian matrix elements are calculated from the
density matrix.

\section{Non-equilibrium Density Matrix}
\label{sec:nedm}

In this section we will first present the relationship between the
scattering state approach and the non-equilibrium Green's function
expression for the non-equilibrium electron density corresponding to
the situation when the electrodes have different electrochemical potentials.
The scattering state approach is quite transparent and has been used
for non-equilibrium first principles calculations by McCann and
Brown,\cite{McBr88} Lang and co-workers,\cite{LaYaIm89,La95,LaAv00}
and Tsukada and co-workers.\cite{HiTs94,HiTs95,KoAoTs01} All these calculations
have been for the case of model jellium electrodes and it is not
straightforward how to extend these methods to the case of electrodes with
a realistic atomic structure and a more complicated electronic 
structure or when localized states are
present inside the contact region.  The use of the non-equilibrium
Green's functions combined with a localized basis set is able to deal
with these points more easily. 

Here we will start with the scattering state approach and make the
connection to the non-equilibrium Green's function expressions for the
density matrix. Consider the scattering states starting in the left
electrode. These are generated from the unperturbed incoming states
(labeled by $l$) of the uncoupled, semi-infinite electrode,
$\psi^0_l$, using the retarded Green's function, $G$, of the coupled
system,
\begin{equation}
\psi_l(\vec x) = 
\psi^0_l(\vec x) + 
\int d\vec{y}\,\, G(\vec x, \vec y)\,V_L(\vec y)\,\psi^0_l(\vec y)\,.
\label{eq:realspaceG}
\end{equation}
As in the
previous section there is no {\em direct} interaction between the
electrodes:
\begin{eqnarray}
V(\vec{r})& = & V_L(\vec{r}) + V_R(\vec{r})\,.\\ 
V_R(\vec{r})\psi^0_l(\vec r)& = & V_L(\vec{r})\psi^0_r(\vec r)=0\,.
\end{eqnarray}

Our non-equilibrium situation is described by the
following scenario: The states starting deep in the
left/right electrode are filled up to the electrochemical potential of the 
left (right) electrode, $\mu_L$ ($\mu_R$). We construct the density matrix
from the (incoming) scattering states of the left and right electrode:

\begin{eqnarray}
D(\vec{x},\vec{y}) &=& \sum_l \psi_l(\vec{x})\psi_l^*(\vec{y})
\,n_F({\varepsilon}_l - {\mu}_L)  \nonumber \\
&+& \sum_r \psi_r(\vec{x})\psi_r^*(\vec{y})
\,n_F(\varepsilon_r-\mu_R)  \,,
\label{eq:realspaceDM}
\end{eqnarray}
where index $l$ and $r$ run over all scattering states in the left and
right electrode, respectively.  Note that this density matrix only
describes states in $C$ which couple to the continuum of electrode
states -- we shall later in Section~\ref{subsec:locstat} 
return to the states localized in $C$.

\subsection{Localized non-orthogonal basis}

Here we will rather consider the density
matrix defined in terms of coefficients of the scattering states with
respect to the given basis (denoted below by Greek subindexes)
\begin{equation}
\psi_l(\vec x) = \sum_{\mu} c_{l\mu} \phi_\mu(\vec x)\,.
\label{eq:expbasis}
\end{equation}
Thus (\ref{eq:realspaceG}) and (\ref{eq:realspaceDM}) read,
\begin{equation}
c_{l\mu} = c^0_{l\mu} + 
\sum_{\nu} 
\left( {\bf G}(z) {\bf V} \right)_{\mu\nu}\,c^0_{l\nu}\,\,,
\,\,\, z=\varepsilon_l + i\delta\,,
\label{eq:coLS}
\end{equation}
\begin{eqnarray}
{\bf D}_{\mu\nu}&=& 
\sum_l c_{l\mu} c_{l\nu}^*\,n_F(\varepsilon_l-\mu_L) \nonumber \\
&+& \sum_r c_{r\mu} c_{r\nu}^*\,n_F(\varepsilon_r-\mu_R)\,.
\end{eqnarray}
The basis is in general non-orthogonal but this will not introduce any
further complications. As for the Hamiltonian, we
assume that the matrix elements of the overlap, ${\bf S}$, are zero
between basis functions in $L$ and $R$.
The overlap is handled by defining  the Green's function matrix ${\bf
G}(z)$ as the inverse of $(z{\bf S - H})$, 
and including the term $-z{\bf S}$ in the perturbation matrix ${\bf V}$.  
To see this we use the following equations,

\begin{eqnarray}
\left[\varepsilon_l{\bf S}_0-{\bf H}_0\right]c^0_{l}  & = &0 \,, \label{eq:ho}\\
\left[\varepsilon_l{\bf S}-{\bf H}\right]c_{l}        & = &0 \,,\label{eq:h}\\
\left[z{\bf S}-{\bf H}\right]{\bf G}(z)& = & \openone         \,.
\end{eqnarray}
With these definitions we see that (\ref{eq:coLS}) is fulfilled,
\begin{equation}
\left[\varepsilon_l{\bf S} - {\bf H}\right]c_{l} = 
\left[\varepsilon_l{\bf S} - {\bf H}\right]c^0_{l} + {\bf V}c^0_{l} = 0 \,,
\end{equation}
when
\begin{equation}
{\bf V}={\bf H}-{\bf H}_0 - \varepsilon_l({\bf S}-{\bf S}_0) \,.
\end{equation}
The use of a non-orthogonal basis is
described further in Refs.~\onlinecite{WiFeLa82} and \onlinecite{EmKi98}.

The density matrix naturally splits into left and right parts. The
derivations for left and right are similar, so we will
concentrate on left only. It is convenient to introduce the left
spectral density matrix, $\rho_L$,
\begin{equation}
{\bf \rho}^L_{\mu\nu}(\varepsilon) =
\sum_l c_{l\mu} c_{l\nu}^*\, \delta(\varepsilon - \varepsilon_l),
\label{eq:rhol}
\end{equation}
and likewise a right spectral matrix $\rho_R$. The density matrix is then
written as,
\begin{equation}
{\bf D}_{\mu\nu}= \int_{-\infty}^{\infty} d\varepsilon\, 
{\bf \rho}^L_{\mu\nu}(\varepsilon)n_F(\varepsilon-\mu_L)+ 
{\bf \rho}^R_{\mu\nu}(\varepsilon)n_F(\varepsilon-\mu_R) \,.
\label{eq:Dspec}
\end{equation}

As always we wish to express ${\bf\rho}^L$ in terms of known
(unperturbed) quantities, i.e., $c^0_{l\mu}$, and for this we use Equation (\ref{eq:coLS}). Since we are only
interested in the density matrix part corresponding to the 
scattering region ($L-C-R$), we note that the coefficients $c^0_{l\mu}$ for the unperturbed 
states are zero for basis functions ($\mu$) within this region. Thus

\begin{equation}
c_{l\mu} = 
\sum_{\nu} 
\left( {\bf G}{\bf V} \right)_{\mu\nu}\,c^0_{l\nu}\,\,,
\end{equation}
where $\nu$ is inside the bulk of the left electrode.
Inserting this in Eq.~(\ref{eq:rhol}) we get,
\begin{equation}
{\bf \rho}^L_{\mu\nu}(\varepsilon)  = 
 \left( 
   {\bf G}(\varepsilon) 
   \frac{1}{\pi}{\mbox{Im}}\left[{\bf V} {\bf g}^{L}(\varepsilon){\bf V}^\dagger\right]
   {\bf G}^\dagger(\varepsilon) \right)_{\mu\nu} .
\end{equation}
Here we use the unperturbed left retarded Green's function,
\begin{equation}
{\bf g}^{L}_{\mu\nu}(\varepsilon) =
\sum_l \frac{c^0_{l\mu} c_{l\nu}^{0*}}
{\varepsilon - \varepsilon_l + i\delta} \,\,,
\end{equation}
and the relation
\begin{equation}
\left[{\bf g}^L(\varepsilon) - 
({\bf g}^L(\varepsilon))^\dagger\right]_{\mu\nu} = 
2\pi i \sum_l  c^0_{l\mu} c_{l\nu}^{0*}\,
\delta(\varepsilon - \varepsilon_l) \,,
\end{equation}
and that ${\bf g}={\bf g}^T$ due to time-reversal symmetry.

We can identify the retarded self-energy,
\begin{eqnarray}
{\bf \Sigma}_L(\varepsilon) &\equiv&
\left[{\bf V}{\bf g}^{L}(\varepsilon)
\,{\bf V}^{\dagger}\right] \,, \\
{\bf \Gamma}_L(z) &\equiv& 
i\left[{\bf \Sigma}_L(\varepsilon) - {\bf \Sigma}_L(\varepsilon)^\dagger\right]/2 \,,
\label{eq:gamma}
\end{eqnarray}
and finally we express ${\bf \rho}^L$ as,
\begin{equation}
{\bf \rho}^L_{\mu\nu}(\varepsilon)  = 
 \frac{1}{\pi}\left( 
   {\bf G}(\varepsilon) 
   {\bf \Gamma}_L(\varepsilon) 
   {\bf G}^\dagger(\varepsilon) \right)_{\mu\nu} \,,
\label{eq:rhoL}
\end{equation}
and a similar expression for ${\bf \rho}^R$.
Note that the ${\bf \Sigma}$, ${\bf \Gamma}$ and ${\bf G}$ matrices in
the equations above are all matrices defined only in the scattering
region $L-C-R$ which is desirable from a practical point of view. 
The {\bf G} matrix is obtained by inverting the matrix in
Eq. \ref{eq:Hamil}.

The expression derived from the scattering states is the same as one
would get from a non-equilibrium Green's function derivation, see e.g.
Refs.~\onlinecite{HaJa96}, where ${\bf D}$ is expressed via the
``lesser'' Green's function,
\begin{equation}
{\bf D} = \frac{1}{2\pi i}\int_{-\infty}^{\infty} d\varepsilon\, 
{\bf G}^<(\varepsilon) \,,
\end{equation}
which includes the information about the non-equilibrium occupation.

\subsection{Complex contour for the equilibrium density matrix}

In equilibrium we can 
combine the left and right parts in Eq.~(\ref{eq:Dspec}), 
\begin{eqnarray}
{\bf G \Gamma G}^{\dagger} &=&  {i \over 2}
 {\bf G} \left[{\bf \Sigma} - {\bf \Sigma}^\dagger\right] {\bf G}^{\dagger}
\nonumber \\
&=& -{i \over 2} 
{\bf G} \left[({\bf G})^{-1} - ({\bf G}^\dagger)^{-1}\right] {\bf G}^{\dagger}
\nonumber \\
&=& -\text{Im}[{\bf G}]
\label{eq:negldelta}
\end{eqnarray}
where ${\bf \Sigma}$ includes both ${\bf \Sigma}_L$ and ${\bf
\Sigma}_R$, and time-reversal symmetry (${\bf G}^{\dagger}= {\bf G}^{*}$)
was invoked.
With this Eq.~(\ref{eq:Dspec}) reduces to the well-known expression,
\begin{eqnarray}
{\bf D}& = & -\frac{1}{\pi}\int_{-\infty}^{\infty} d\varepsilon\,\,
\text{Im}[{\bf G}(\varepsilon+i\delta)]\, n_F(\varepsilon-\mu)
\nonumber \\ & = & -\frac{1}{\pi}\text{Im}\Big[\int_{-\infty}^{\infty}
d\varepsilon\, {\bf G}(\varepsilon+i\delta)\,
n_F(\varepsilon-\mu)\Big]                           \,.
\label{eq:Grealint}
\end{eqnarray}
The invocation of time-reversal symmetry 
makes ${\bf D}$ a real matrix since ${\bf D}^*={\bf D}^T={\bf D}$. 

At this point it is important to note that we have neglected the
infinitesimal $i\delta$ in Eq.~(\ref{eq:negldelta}). This means that
the equality in Eq.~(\ref{eq:negldelta}) is actually not true when
there are states present in $C$ which do not couple to any of the
electrodes, and thus ${\bf \Gamma}_L={\bf\Gamma}_R=0$ and
$\rho_L=\rho_R=0$ for elements involving strictly localized states.
The localized states cannot be reached starting from scattering states
and are therefore not included in Eq.~(\ref{eq:Dspec}), while they are
present in Eq.~(\ref{eq:Grealint}). We return to this point in 
Section~\ref{subsec:locstat} below.

\begin{figure}[thb]
  \epsfig{figure=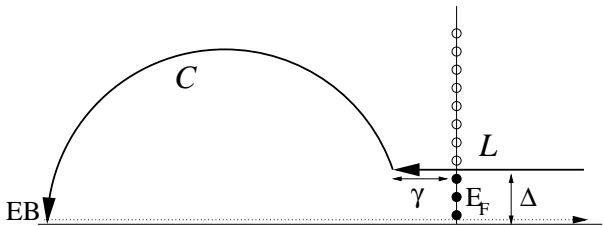, width=0.95\columnwidth, angle=0}
\caption{The closed contour: $L$ ($]\infty+i\Delta;EF-\gamma+i\Delta[$), $C$, 
and $[EB+i\delta;\infty+i\delta]$ enclosing the Fermi poles (black dots). }
\label{fig:contour}
\end{figure}

All poles of the retarded Green's function ${\bf G}(z)$ are
lying on the real axis and the function is analytic otherwise.
Instead of doing the integral in Eq.~(\ref{eq:Grealint}) (corresponding to
the dotted line in Fig.~\ref{fig:contour}), we consider the contour in
the complex plane defined for a given finite temperature shown by the 
solid line in
Fig.~\ref{fig:contour}. Indeed, the closed contour beginning with line
segment $L$, followed by the circle segment $C$, and running along the
real axis from $(EB+i\delta)$ to $(\infty+i\delta)$, where $EB$ is  below 
the bottom valence band edge, will only enclose
the poles of $n_F(z)$ located at $z_\nu=i(2\nu+1)\pi kT$. According
to the residue theorem,
\begin{equation}
\oint dz\, {\bf G}(z)\, n_F(z-\mu) =  -2\pi i\, kT \sum_{z_\nu} {\bf
G}(z_\nu) \,,
\end{equation}
where we use that the residues of $n_F$ are $-kT$. Thus,
\begin{eqnarray}
&\int_{EB}^{\infty} d\varepsilon\, 
{\bf G}(\varepsilon+i\delta)\, n_F(\varepsilon-\mu)
\label{eq:Dcontint}\\
& = -\int_{C+L}dz\,{\bf G}(z)\, n_F(z-\mu) -
2\pi i\, kT \sum_{z_\nu} {\bf G}(z_\nu) \,. \nonumber
\end{eqnarray}

\begin{figure}[thb]
  \epsfig{figure=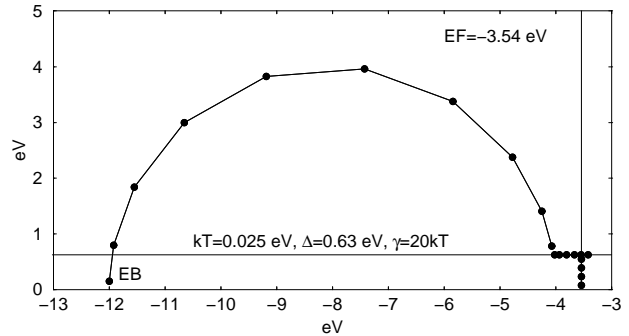, width=0.95\columnwidth, angle=0}
\caption{Typical points for Gaussian quadrature on the contour. On $L$ 
we employ a quadrature with a weight function equal to the
Fermi function.}
\label{fig:gauss}
\end{figure}

The contour integral can be computed numerically for a given finite
temperature by choosing the number of Fermi poles to enclose. This
insures that the complex contour stays away from the real axis (the
part close to $EB$ is not important). The Green's function will behave
smoothly sufficiently away from the real axis, and we can do the
contour integral by Gaussian quadrature with just a minimum number of
points, see Fig.~\ref{fig:gauss}. The main variation on $L$ comes from
$n_F$ and it is advantageous to use $n_F$ as a weightfunction in the
Gaussian quadrature.\cite{gaussfermi}

\subsection{Numerical procedure for obtaining the non-equilibrium density 
matrix}

In non-equilibrium the density matrix is given by
\begin{eqnarray}
{\bf D}_{\mu\nu}&= &{\bf D}^L_{\mu\nu} + \Delta^R_{\mu\nu}\,,  \label{eq:D1}\\
{\bf D}_{\mu\nu}^L & = & -\frac{1}{\pi}\text{Im}\Big[\int_{EB}^{\infty} d\varepsilon\, 
{\bf G}(\varepsilon+i\delta)\, n_F(\varepsilon-\mu_L)\Big] \,, \label{eq:D1eq}\\
\Delta^R_{\mu\nu} & =& \int_{-\infty}^{\infty} d\varepsilon\, 
{\bf \rho}^R_{\mu\nu}(\varepsilon)
[
n_F(\varepsilon-\mu_R) - n_F(\varepsilon-\mu_L)
]\,,
\end{eqnarray}
or equivalently
\begin{eqnarray}
{\bf D}_{\mu\nu}&= &{\bf D}^R_{\mu\nu} + \Delta^L_{\mu\nu} \,, \label{eq:D2}\\
{\bf D}_{\mu\nu}^R & = & -\frac{1}{\pi}
\text{Im}\Big[\int_{EB}^{\infty} d\varepsilon\, 
{\bf G}(\varepsilon+i\delta)\, n_F(\varepsilon-\mu_R)\Big] \,,  \label{eq:D2eq}\\
\Delta^L_{\mu,\nu} & =& \int_{-\infty}^{\infty} d\varepsilon\, 
{\bf \rho}^L_{\mu\nu}(\varepsilon)
[
n_F(\varepsilon-\mu_L) - n_F(\varepsilon-\mu_R)
]                             \,.
\end{eqnarray}

The spectral density matrices, ${\bf \rho}^L$ and ${\bf \rho}^R$, are
not analytical. Thus only the ``equilibrium'' part of the density
matrix, ${\bf D}^L$(${\bf D}^R$), can be obtained using the complex
contour. Furthermore, this is a real quantity due to the time-reversal
symmetry, whereas the ``non-equilibrium'' part, $\Delta^L$($\Delta^R$)
cannot be made real since the scattering states by construction break
time-reversal symmetry due to their boundary conditions. The imaginary
part of $\Delta^L$ ($\Delta^R$) is in fact directly related to the
local current.\cite{Todorov99} However, if we are interested only in
the electron density and if we employ a basis-set with real basis
functions ($\phi_\mu$) we can neglect the imaginary part of ${\bf D}$,
\begin{equation}
n(\vec r)=\sum_{\mu,\nu} \phi_\mu(\vec r) 
{\text{Re}}\Big[{\bf D}_{\mu\nu}\Big] \phi_\nu(\vec r) \,.
\end{equation}

To obtain $\Delta^L$ ($\Delta^R$) the integral must be evaluated for a
finite level broadening, $i\delta$, and on a fine grid.  Even for
small voltages we find that this integral can be problematic, and care
must be taken to ensure convergence in the level broadening and number
of grid points. Since we have two similar expressions for the density
matrix we can get the integration error from
\begin{equation}
{\bf e}_{\mu\nu}= {\bf D}^R_{\mu\nu} + 
\Delta^L_{\mu\nu}-({\bf D}^L_{\mu\nu} + \Delta^R_{\mu\nu}).
\label{eq:err}
\end{equation}
The integration error arises mainly from the real axis integrals, and
depending on which entry of the density matrix we are considering
either $\Delta^L$ or $\Delta^R$ can dominate the error.  Thus with
respect to the numerical implementation the two formulas
Eq.~(\ref{eq:D1}) and (\ref{eq:D2}) are not equivalent. We will
calculate the density matrix as a weighted sum of the two integrals
\begin{eqnarray}
\label{eq:wq1}
{\bf D}_{\mu\nu}&= & w_{\mu\nu} 
({\bf D}^L_{\mu\nu} + \Delta^R_{\mu\nu}) + 
(1-w_{\mu\nu})({\bf D}^R_{\mu\nu} + \Delta^L_{\mu\nu})  \\
w_{\mu\nu} & = & 
\frac{(\Delta^L_{\mu\nu})^2}
{(\Delta^L_{\mu\nu})^2 + (\Delta^R_{\mu\nu})^2}.
\label{eq:w2}
\end{eqnarray}
The choice of weights can be rationalized by the following argument.
Assume that the result of the numerical integration is given by a
stochastic variable $\tilde{\Delta}^L$ with mean value $\Delta^L$ and
the standard deviation is proportional to the overall size of the
integral, i.e. $Var(\tilde{\Delta}^L) \propto (\Delta^L)^2$. A numerical
calculation with weighted integrals as in Eq.~(\ref{eq:wq1}) will then
be a stochastic variable with the variance
\begin{equation}
Var(\tilde{\bf D})  \propto w^2( \Delta^R)^2 +  (1-w)^2 ( \Delta^L)^2.
\end{equation}
The value of $w$ which minimize the variance is the weight factor we
use in Eq.~(\ref{eq:w2}). 


\subsection{Localized states}
\label{subsec:locstat}
As mentioned earlier the signature of a localized state at
$\varepsilon_0$ in the scattering region is that the matrix elements
of ${\bf \Gamma}_L(\varepsilon_0),{\bf \Gamma}_R(\varepsilon_0)$ are
zero for that particular state. Localized states most commonly arise
when the atoms in $C$ have energy levels below the bandwidth of the
leads. The localized states give rise to a pole at $\varepsilon_0$ in
the Green's function. As long as $\varepsilon_0 < \{\mu_L, \mu_R\}$ the
pole will be enclosed in the complex contours and therefore included
in the occupied states. If on the other hand the bound state has an
energy within the bias window, i.e. $ \mu_L< \varepsilon_0 < \mu_R$
the bound state will not be included in the real axis integral
($\Delta^L$, $\Delta^R$) and in the complex contour for ${\bf D}^L$,
but it will be included in the complex contour for ${\bf D}^R$. Such a
bound state will only be correctly described by the present formalism
if additional information on its filling is supplied. These
situations are rare and seldom encountered in practice.



\section{The nonequilibrium effective potential}
\label{sec:veff}
The DFT effective potential consists of three parts: a pseudo potential
$V_{ps}$, the exchange correlation potential $V_{xc}$, and the Hartree
potential $V_{H}$.  For $V_{ps}$ we use norm conserving
Troullier-Martins pseudopotentials, determined from standard
procedures.\cite{TrMa91} For $V_{xc}$ we use the LDA as
parametrized in Ref.~\onlinecite{PeZu81}.

\subsection{The Hartree potential}

The Hartree potential is a non-local function of the electron density,
and it is determined through the Poisson's equation (in Hartree atomic units),
\begin{equation}
{\nabla}^{2} V_{H}(\vec{r})=-4 \pi n(\vec{r}) \,.
\end{equation}
Specifying the electron density only in the C region of
Fig.~\ref{fig:setup1D} 
makes the Hartree potential of this region
undetermined up to a linear term\cite{notepoisson},
\begin{equation}
V_{H}(\vec{r})=\phi (\vec{r})+\vec{a}\cdot \vec{r}+b \,,
\label{eq:vh}
\end{equation}
where $\phi (\vec{r})$ is a solution to Poisson's equation in region C
and $\vec{a}$ , $b$ are parameters that must be determined from the
boundary conditions to the Poisson's equation. In the directions
perpendicular to the transport direction ($x$,$y$) we will use
periodic boundary conditions which fix the values of $a_{x}$, $a_{y}$.
The remaining two parameters $a_{z}$, $b$ are determined by the value
of the electrostatic potential at the $L-C$ and $C-R$ boundaries. The
electrostatic potential in the $L$ and $R$ regions could be determined from
the separate bulk calculations, and 
shifted relative to each other by the bias $V$.
With these boundary
conditions the Hartree potential in the contact
is uniquely defined, and  could be
computed using a real-space technique \cite{realspace} or an iterative
method.\cite{HiTs95} 

However, in the present work, we have solved the Poisson's equation
using a Fast Fourier Transform (FFT) technique.  We set up a supercell
with the $L-C-R$ region, which can contain some extra layers of buffer
bulk atoms and, possibly, vacuum (specially if the two electrodes are
not of the same nature, otherwise the $L$ and $R$ are periodically
matched in the $z$ direction). We note in passing that this is done so
that the potential at the $L-C$ and $C-R$ boundaries reproduces the
bulk values, crucial for our method to be consistent.  For a given
bias, $V$, the $L$ and $R$ electrode electrostatic potentials need to
be shifted by $V/2$ and $-V/2$, respectively, and $V_H$ will therefore
have a discontinuity at the cell boundary.  The electrostatic
potential of the supercell is now decomposed as:
\begin{equation}
V_{H}(\vec{r})=\widetilde{\phi }(\vec{r}) - V(\frac{z}{L_z}-0.5).
\end{equation}
where $\widetilde{\phi }(r)$ is a periodic solution of the
Poisson's equation in the supercell, and therefore can be
obtained using FFT's.\cite{notepoisson2}  

\begin{figure}
  \epsfig{figure=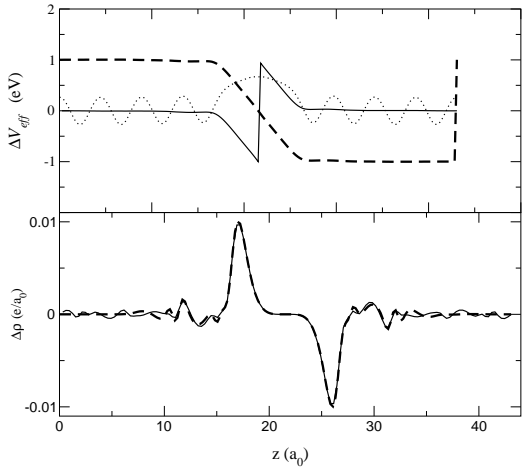, width=0.8\columnwidth, angle=0}
\caption{(a) The induced external potential for slab calculation (full line), 
  and in the {\sc TranSiesta} calculation (dashed line). In the slab
  calculation the jump in external potential is in the middle of the
  vacuum region.  The total potential (arbitrary units) is shown for
  reference (dotted line). (b) Induced density. Potential and density is averaged
  in the surface plane. The density corresponds to one surface unit
  cell. }
\label{fig:surfacefield}
\end{figure}

To test the method we have calculated the induced density and
potential on a ``capacitor'' consisting of two gold (111) surfaces
separated by a 12 bohr wide tunnelgap and with a voltage drop of 2
Volt.  We have calculated the charge density and the potential in this
system in two different ways. First, we apply the present formulation
(implemented in {\sc TranSiesta}), where the system consists of two
semi-infinite gold electrodes, and the Hartree potential is computed
as described above.  Then, we calculate a similar system, but with a
slab geometry, computing the Hartree potential with {\sc Siesta},
adding the external potential as a ramp with a discontinuity in the
vacuum region.  Fig.~\ref{fig:surfacefield} shows the comparison of
the results for the average induced density and potential along the
$z$ axis.  Since the tunnel gap is so wide that there is no current
running, the two methods should give very similar results, as we
indeed can observe in the Figure. We can also observe that the
potential ramp is very effectively screened inside the material, so
that the potential is essentially equal to the bulk one, except for
the surface layer.  This justifies our approach for the partition of
the system, the solution of the Poisson's equation, and the use of the
bulk Hamiltonian matrix elements fo the $L$ and $R$ regions (see
below).

\vspace{0.5truecm}

\subsection{The Hamiltonian matrix elements}
Having determined the effective potential we calculate the Hamiltonian
matrix elements as in standard {\sc Siesta} calculations.  However,
since we only require the density and the electrostatic potential to
be correct at the $L-C$ and $C-R$ boundaries, the ${\bf H}_L$ and
${\bf H}_R$ parts of the Hamiltonian (see Eq.~(\ref{eq:Hamil})) will
not be correct.  We therefore substitute ${\bf H}_L$ and ${\bf H}_R$
with the Hamiltonian obtained from the calculation of the separate
bulk electrode systems. Here it is important to note that the
effective potential within the bulk electrode calculations usually are
shifted rigidly relative to the effective potential in the $L$ and $R$
regions, due to the choice of the parameter $b$ in Eq.~(\ref{eq:vh}).
However, the bulk electrode Hamiltonians ${\bf H}_L$ and ${\bf H}_R$
can easily be shifted, using the fact that the electrode Fermi level
should be similar to the Fermi level of the initial {\sc Siesta}
calculation for the BLCRB supercell.

The discontinuity of the Hartree potential at the cell boundary has no
consequence in the calculation: the Hamiltonian matrix elements
{\em inside} the $L-C-R$ region are unaffected because of the
finite range of the atomic orbitals, and the Hamiltonian matrix
elements {\em outside} the $L-C-R$ region which do feel the discontinuity
are replaced by bulk values (shifted according to the bias).


\section{Conductance formulas}
\label{sec:conduc}

Using the non-equilibrium Green's function formalism 
(see e.g. Refs.~\onlinecite{HaJa96,Datta,BrKoTs99} and references therein) 
the current, $I$, through the contact can be derived
\begin{eqnarray}
I(V)= G_0&& 
 \int_{-\infty}^{\infty}d\epsilon\, 
 \left(n_F(\epsilon - \mu_L)-n_F(\epsilon - \mu_R)\right)\nonumber \\
&&{\text{Tr}}
 \left[ 
{\bf \Gamma}_L(\epsilon)
 {\bf G}^\dagger(\epsilon)
{\bf \Gamma}_R(\epsilon)
 {\bf G}(\epsilon)     
 \right] \,,
\label{eq:orthcond}
\end{eqnarray}
where $G_0=2e^2/h$. We note that this expression is not general but is
valid for mean-field theory like DFT.\cite{MeWi92} An equivalent
formula has been derived by Todorov {\it et al.}  \cite{ToBrSu93} (the
equivalence can be derived using Eq.~(\ref{eq:gamma}) and the cyclic
invariance of the trace in Eq.~\ref{eq:orthcond}).

With the identification of the (left-to-right) transmission 
amplitude matrix ${\bf t}$,\cite{CuYeMR98}
\begin{equation}
{\bf t}(\epsilon)= \left(
{\bf \Gamma}_R(\epsilon)\right)^{1/2}
{\bf G}(\epsilon)
\left(
{\bf \Gamma}_L(\epsilon)\right)^{1/2}   \,,
\label{eq:tfromg}
\end{equation}
(\ref{eq:orthcond}) is seen to be equivalent to the
Landauer-B{\"u}ttiker formula\cite{BuImLa85} for the conductance, $G=I/V$,
\begin{eqnarray}
G(V)&=&\frac{G_0}{V}
 \int_{-\infty}^{\infty}d\varepsilon\, 
 [ n_F(\varepsilon - \mu_L)-n_F(\varepsilon - \mu_R) ]\nonumber\\
 &&{\text{Tr}}\left[{\bf t}^\dagger{\bf t}\right](\varepsilon) \,,
\label{eq:landauer}
\end{eqnarray}
The eigenchannels are defined in terms of the (left-to-right)
transmission matrix ${\bf t}$,\cite{Buttike88a,MaLa92}
\begin{equation}
{\bf t}\,=\,{\bf U}_R\, \mbox{diag}\{\vert\tau_n\vert\}\, 
{\bf U}_L^\dagger\, ,
\label{eq:utu}
\end{equation} 
and split the total transmission into individual channel contributions,
\begin{equation}
T_{\text{Tot}} = \sum_n\,\vert\tau_n\vert^2 \,.
\label{eq:diagland}
\end{equation} 
The collection of the individual channel transmissions
$\{ \vert\tau_n\vert^2 \}$ gives a more detailed description of the
conductance and is useful for the interpretation of the 
results.\cite{BrSoJa97,CuYeMR98,BrKoTs99}


\section{Applications}


\subsection{Carbon wires/Aluminum (100) electrodes}

Short monoatomic carbon wires coupled to metallic electrodes have
recently been studied by Lang and Avouris\cite{LaAv98,LaAv00} and
Larade {\it et al.}~\cite{LaTaMeGu01}.  Lang and Avouris used the
Jellium approximation for the electrodes, while Larade~{\it et al.}
used Al electrodes with a finite cross section oriented along the
(100) direction.  In this section, we will compare the {\sc
  TranSiesta} method with these other first principles electron
transport methods by studying the transmission through a 7-atom carbon
chain coupled to Al(100) electrodes with finite cross sections as well
as to the full Al(100) surface.

\begin{figure}
\epsfig{figure=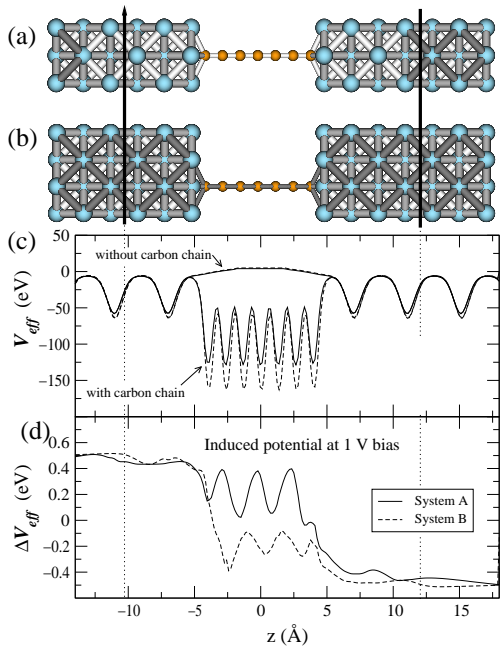, width=0.8\columnwidth, angle=0}
\caption{(a) The 7-atom carbon chain with finite cross section Al(100)
  electrodes(system A). (b) The carbon chain with
  Al(100)-($2\sqrt{2}\times2\sqrt{2}$) electrodes(system B). (c) The
  effective potential of system A(dashed) and system B(solid),
  together with the effective potential of the corresponding bare
  electrode systems.  (d) The selfconsistent effective
  potential for an external bias of 1 V (the zero bias effective
  potential has been subtracted).\label{fig:Cwire} }
\end{figure}

We consider two systems, denoted A and B, shown in
Fig.~\ref{fig:Cwire}a,b. System A consists of a 7-atom carbon chain
coupled to two electrodes of finite cross section oriented along the
Al(100) direction (see Fig.~\ref{fig:Cwire}a). The electrode unit cell
consists of 9 Al atoms repeated to $z = \pm \infty $. The ends of the
carbon chain are positioned in the Al(100) hollow site and the
distance between the ends of the carbon chain and the first plane of
Al atoms is fixed to be $d = 1.0$ \AA . In system B the carbon chain
is coupled to two Al(100)-($2\sqrt{2}\times2\sqrt{2}$) surfaces with
an Al-C coupling similar to system A. In this case the electrode unit
cell contains 2 layers each with 8 atoms. For both systems the contact region ($C$)
includes 3 layers of atoms in the left electrode and 4 layers of the
right electrode. We use single-$\zeta$ basis sets for both C and Al
to be able to compare with the results from McDCAL, which were
obtained with that basis.\cite{LaTaMeGu01}

The conductance of system A is dominated by the alignment of the LUMO
state of the isolated chain to the Fermi level of the electrodes
through charge transfer.\cite{LaTaMeGu01} The coupling of the LUMO,
charge transfer, and total conductance can be varied continuously by
adjusting the electrode-chain separation.\cite{LaTaMeGu01} For our
value of the electrode-chain separation we get a charge transfer of
1.43 and 1.28 $e$ to the carbon wire in systems A and B, respectively.
This is slightly larger than the values obtained by Lang and
Avouris\cite{LaAv00} for Jellium electrodes, but in good agreement
with results from McDCAL.\cite{LaTaMeGu01}

\begin{figure}
 \epsfig{figure=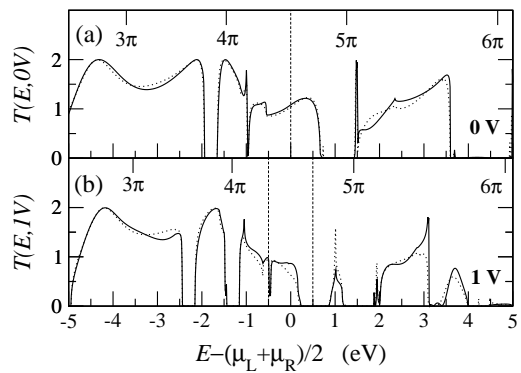, width=0.8\columnwidth, angle=0}
\caption{(a) Zero bias transmission coefficient, $T(E,0V)$,  
  for the 7-atom carbon chain with finite cross section Al(100)
  electrodes (system A).  (b) Transmission coefficient at 1 V,
  $T(E,1V)$. Solid lines show results obtained with {\sc TranSiesta},
  and dotted lines results obtained with McDCAL.  The vertical dashed
  lines indicated the window between $\mu_L$ and $\mu_R$.  The
  position of the eigenstates of the carbon wire subsystem are also
  indicated at the top axis.
\label{fig:Cwirelead}}
\end{figure}

To facilitate a more direct comparison between the methods we show in
Fig.~\ref{fig:Cwirelead} the transmission coefficient of system A
calculated both within {\sc TranSiesta} (solid) and McDCAL (dotted).
For both methods, we have used identical basis sets and
pseudopotentials.  However, several technical details in the
implementations differ and may lead to small differences in the
transmission spectra.  The main implementation differences between the
two methods are related to the calculation of Hamiltonian parameters
for the electrode region, the solution of the Poisson's equation, and
the complex contours used to obtain the electron
charge.\cite{trans-mcdcal} Thus, there are many technical differences
in the two methods, and we therefore find the close agreement in
Fig.~\ref{fig:Cwirelead} very satisfactory.

\begin{figure}
 \epsfig{figure=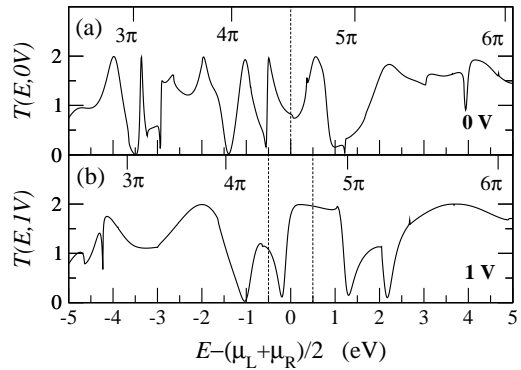, width=0.8\columnwidth, angle=0}
\caption{ Transmission coefficients,  $T(E,0V)$ and  $T(E,1V)$  for the 7-atom carbon chain
  with Al(100)-($2\sqrt{2}\times2\sqrt{2}$) electrodes(system B).  The
  position of the eigenstates of the carbon wire subsystem are also
  indicated at the top axis.
\label{fig:Cwire100}}
\end{figure}

In Fig.~\ref{fig:Cwire100} we show the corresponding transmission
coefficients for system B.  It can be seen that the transmission
coefficient for zero bias at $\varepsilon=\mu$ is close to 1 for both
systems, thus they have similar conductance. However, the details in
the transmission spectra differs much from system A. In order to get
some insight into the origin of the different features we have
projected the selfconsistent Hamiltonian onto the carbon orbitals, and
diagonalised this subspace Hamiltonian to find the position of the
carbon eigenstates in the presence of the Al electrodes.  Within the
energy window shown in Fig.~\ref{fig:Cwirelead} and \ref{fig:Cwire100}
we find 4 doubly degenerate $\pi$-states (3$\pi$, 4$\pi$, 5$\pi$,
6$\pi$). The positions of the eigenstates are indicated above the
transmission curves. Each doubly degenerate state can contribute to
the transmission with 2 at most. Generally, the position of the carbon
$\pi$ states give rise to a slow variation in the transmission
coefficient, and the fast variation is related to the coupling between
different scattering states in the electrodes and the carbon
$\pi$-states. For instance in system A, there are two energy intervals
[-1.9,-1.7] and [0.7,1.4], where the transmission coefficient is zero,
and the scattering states in these energy intervals are therefore not
coupling to the carbon wire. Note how these zero transmission
intervals are doubled at finite bias, since the scattering states of
the left and right electrode are now displaced.
 
The energy dependence of the transmission coefficient is quite
different in system B compared to system A. This is mainly due to the
differences in the electronic structure of the surface compared to the
finite sized electrode.  However, the electronic states of the carbon
wires are also slightly different. We find that the $\pi$-states lie
$~0.2$ eV higher in energy in system B relative to system A.  As
mentioned previously, the charge transfer to the carbon chain is
different in the two systems. The origin of this is related to a
larger work function ($\sim 1 $eV) of the surface relative to the
lead. We note that the calculated work function of the surface is 0.76
eV higher than the experimental workfunction of Al (4.4
eV),\cite{CRC94} which may be due to the use of a single-$\zeta$ basis
set and the approximate exchange-correlation description.\cite{SkRo92}
 
In Fig.~\ref{fig:Cwire}d, we show the changes in the effective
potential when a 1 V bias is applied.  We find that the potential does
not drop continuously across the wire.  In system A, the main
potential drop is at the interface between the carbon wire and the
right electrode, while in system B the potential drop takes place at
the interface to the left electrode. This should be compared to the
Jellium results, where there is a more continuous voltage drop through
the system.\cite{LaAv00} We do not yet understand the details of the
origin of these voltage drops. However, it seems that the voltage drop
is very sensitive to the electronic structure of the electrodes.
Thus, we find it is qualitatively and quantitatively important to have
a good description of the electronic structure of the electrodes.

\subsection{Gold wires/Gold (111) electrodes}
The conductance of single atom gold wires is a benchmark in atomic
scale conduction. Since the first
experiments\cite{AgRoVi93,PaMeGo93,OlLaSt94} numerous detailed studies
of their conductance have been carried out through the 1990s until now
(see e.g. Ref.~\onlinecite{Ruiten00} for a review). More recently the
non-linear conductance\cite{YaSa97,CoCoZh99,HaNiBr00,YuEnSa01} has
been investigated and the atomic
structure\cite{OhKoTa98,RoFuUg00,RoUg01} of these systems has been
elucidated. Experiments show that chains containing more than 5 gold
atoms\cite{YaBoBr98} can be pulled and that these can remain stable
for an extended period of time at low temperature. A large number of
experiments employing different techniques and under a variety of
conditions (ambient pressure and UHV, room and liquid He temperature) all 
show that the conductance at low bias is very close to 1 $G_0$ and several
experiments point to the fact that this is due to a single conductance
eigenchannel.\cite{ScAgCu98,BrRu99,LuDeEs99}

Several theoretical investigations have addressed the stability and
morphology~\cite{ToToCoErKoToSo99,SaArJu99,OkTa99,HaBaLa99,MaSp00,HaBaSc00}
and the conductance~\cite{OkTa99,HaBaLa99,HaBaSc00} of atomic gold chains and
contacts using DFT. However, for the
evaluation of the conductance, these studies have neglected the
presence of valence $d$-electrons and the scattering due to the
non-local pseudopotential. This approximation is not justified 
{\em a priori}: for example, it is clear that the bands due to
$d$-states are very close to the Fermi level in infinite linear chains
of gold and this indicates that these could play a role, especially for a
finite bias.\cite{BrKoTs99} 

\subsubsection{Model}

In this section we consider gold wires connected between the (111)
planes of two semi-infinite gold electrodes. In order to keep the
computational effort to a minimum we will limit our model of the
electrode system to a small unit cell ($3\times 3$) and use only the
$\Gamma$-point in the transverse (surface) directions. We have used a
single-$\zeta$ plus polarization basis set of 9
orbitals corresponding to the $5d$ and $6(s,p)$ states of the free
atom. In one calculation (the wire labelled (c) in
Fig.~\ref{fig:3struc}) we used double-$\zeta$ representation of the
$6s$ state as a check and found no significant change in the results.
The range of interaction between orbitals is limited by the radii
of the atomic orbitals to 5.8 {\AA},
corresponding to the 4th nearest neighbor in the bulk gold crystal or
a range of three consecutive layers in the [111] direction.  We have
checked that the bandstructure of bulk gold with this basis set is in
good agreement with that obtained with more accurate basis sets for
the occupied and lowest unoccupied bands.

\begin{figure}
  \epsfig{figure=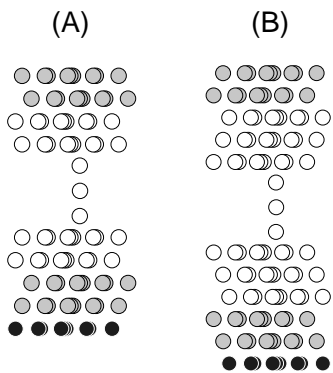, width=0.5\columnwidth, angle=0}
\caption{Models used for the gold wires calculations.
  The white atoms correspond to contact region $C$ while the
  grey atoms correspond to the $L$ and $R$ regions in Fig.~\ref{fig:setup1D}.
  The black atoms are only included in the initial {\sc Siesta}
  calculation and can be added in order to yield a better initial
  density matrix for the subsequent {\sc TranSiesta} run. We have used 2(A)
  and 3(B) surface layers in the contact region.}
\label{fig:goldsetup}
\end{figure}

We have considered two different configurations of our calculation
cell, shown in Fig.~\ref{fig:goldsetup}. In most calculations we
include two surface layers in the contact region ($C$) where the
electron density matrix is free to relax and we have checked that
these results do not change significantly when three surface layers
are included on both electrodes.  
We obtain the initial guess for a
density matrix at zero bias voltage from an initial {\sc Siesta} calculation
with normal periodic boundary conditions in the transport ($z$)
direction.\cite{siestaprecalc} In order to make this density matrix as close to the
{\sc TranSiesta} density matrix we can include extra layers in the interface
between the $L$ and $R$ regions (black atoms in
Fig.~\ref{fig:goldsetup}) to simulate bulk. In the case of two
different materials for $L$ and $R$ electrodes many layers may be
needed, but in this case we use just one layer. We use the zero bias
{\sc TranSiesta} density matrix as a starting point for 
{\sc TranSiesta} runs with finite bias.

\subsubsection{Bent wires}

In a previous study by S\'anchez-Portal and co-workers,\cite{SaArJu99}
a zig-zag arrangement of the atoms was found to be energetically
preferred over a linear structure in the case of infinite atomic gold
chains, free standing clusters, and short wires suspended between two
pyramidal tips. In general the structure of the wires will be
determined by the fixed distance between the electrodes and the wires
will therefore most probably be somewhat compressed or stretched.

\begin{figure}
  \epsfig{figure=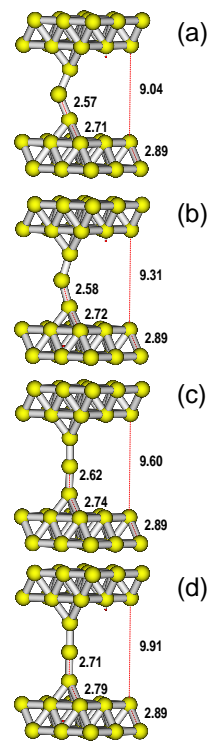, width=0.3\columnwidth, angle=0}
\caption{We have considered the distances 
  9.0 (a), 9.3 (b), 9.6 (c) and 9.9 (d) {\AA} between the two (111)
  surfaces. All wires have been relaxed while the surface atoms are
  kept fixed. Distances are shown in {\AA}.}
\label{fig:3struc}
\end{figure}

Here we have considered wires with a length of three atoms and
situated between the (111) electrodes with different spacing.
Initially the wire atoms are relaxed at zero voltage bias (until any
force is smaller than 0.02 eV/\AA) and for fixed electrode atoms. The
four relaxed wires for different electrode spacings
are shown in Fig.~\ref{fig:3struc}(a)-(d). The
values for bondlength and bondangle of the first wire (a),
$r=2.57$~\AA, $\alpha=135^\circ$, are close to the values found in
Ref.~\onlinecite{SaArJu99} for the infinite periodic wires at the
minimum of energy with respect to unit-cell length ($r=2.55$ ~\AA,
$\alpha=131^\circ$).

\begin{figure}
  \epsfig{figure=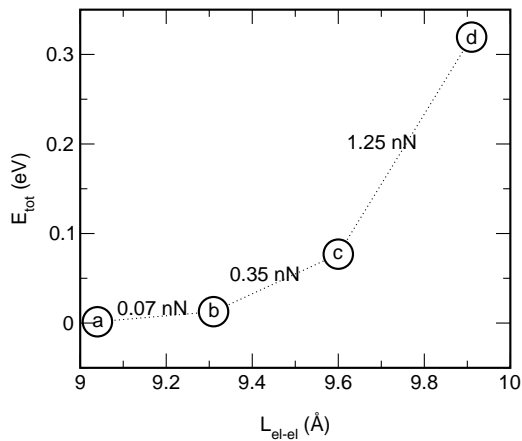, width=0.8\columnwidth, angle=0}
\caption{The change in total energy of the relaxed wires 
  during the elongation shown in Fig.~\ref{fig:3struc} as calculated
  in a {\sc Siesta} run. The force determined from the slopes of the line
  segments is shown also.}
\label{fig:etotforce}
\end{figure}

\begin{figure}
  \epsfig{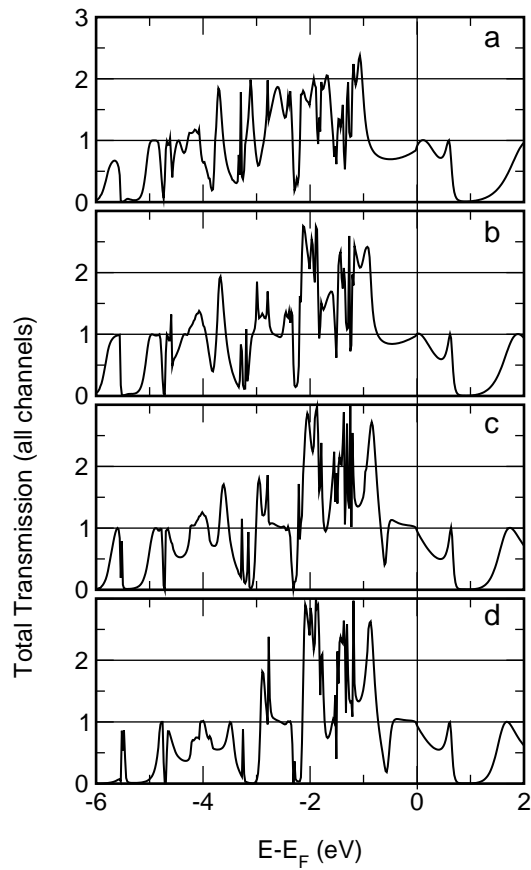}
\caption{The total transmission of the wires shown in Fig.~\ref{fig:3struc} vs. electron energy.}
\label{fig:tottrans}
\end{figure}

In Fig.~\ref{fig:etotforce} we show the total energy and corresponding
force as evaluated in a standard {\sc Siesta} calculation for the
wires as a function of electrode spacing. The force just before the
stretched wire breaks has been measured\cite{RuAgVi96,RuBaAg01} and is
found to be 1.5$\pm$0.3 nN independent of chain length. The total
transmission resolved in energy is shown in Fig.~\ref{fig:tottrans}
for zero bias.  The conductance in units of $G_0$ is given by
$T_{\text{Tot}}(E_F)$ which is $0.91$, $1.00$, $0.95$, and $0.94$ for
the (a), (b), (c), and (d) structures of Fig.~\ref{fig:3struc},
respectively. It is striking that the measured conductance in general
stays quite constant as the wire is being stretched. Small dips below
1 $G_0$ can be seen, which might be due to additional atoms being
introduced into the wire from the electrodes during the
pull.\cite{RuBaAg01} It is interesting to note that the value for (a)
is quite close to the conductance dip observed in
Ref.~\onlinecite{RuBaAg01} and we speculate that this might correspond
to the addition of an extra atom in the chain which will then attain a
zig-zag structure which is subsequently stretched out to a linear
configuration.

\begin{figure}
  \epsfig{figure=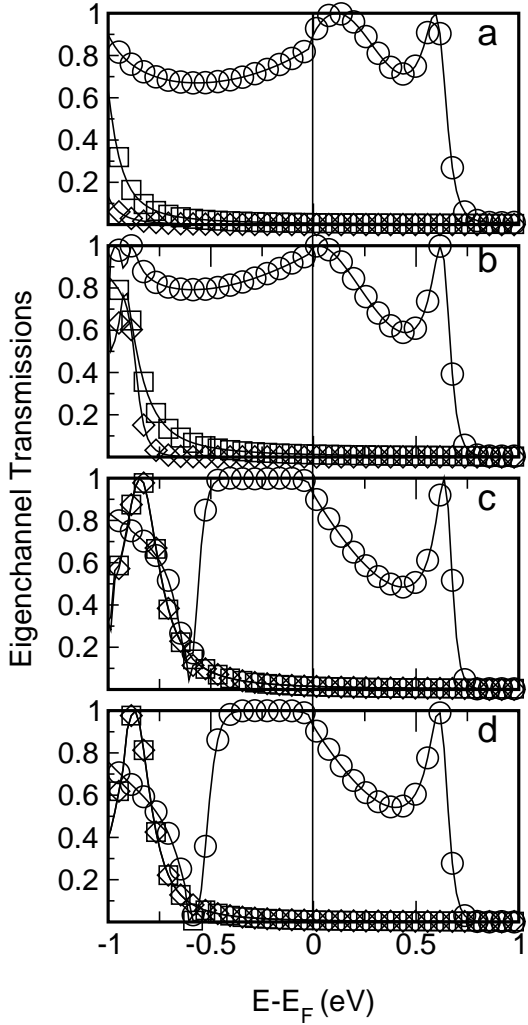, width=0.8\columnwidth, angle=0}
\caption{Eigenchannel transmissions of the wires in Fig.~\ref{fig:3struc}. Only 
three channels give significant contribution within the energy range shown.}
\label{fig:allteig}
\end{figure}

It can be seen from the corresponding eigenchannel decomposition in
Fig.~\ref{fig:allteig} that the conductance is due to a single, 
highly transmitting channel, in agreement with the experiments
mentioned earlier and previous
calculations.\cite{BrSoJa97,CuYeMR98,BrKoTs99,HaBaSc00} This channel
is composed of the $l_z=0$ orbitals.\cite{BrKoTs99} About $0.5-1.0$ eV
below the Fermi energy transmission through additional channels is
seen. These are mainly derived from the $l_z=1$ orbitals and are
degenerate for the wires without a bend, due to the rotational symmetry.

\begin{figure}
  \epsfig{figure=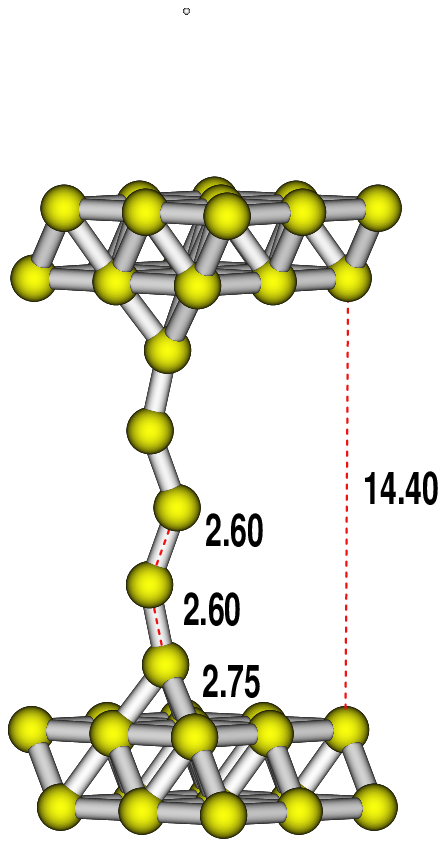, width=0.3\columnwidth, angle=0}
\caption{Relaxed structure of a 5 atom long chain. Distances are shown in {\AA}. }
\label{fig:5struc}
\end{figure}

We have done a calculation for a 5 atom long wire. The relaxed
structure is shown in Fig.~\ref{fig:5struc}. We note that while the
bondlengths are the same within the wire, there is a different
bondangle ($143^\circ$ in the middle, $150^\circ$ at the electrodes).
We find that the transmission at zero bias is even closer to unity
compared with the 3 atom case and find a conductance of 0.99 $G_0$
despite its zig-zag structure.

\begin{figure}
  \epsfig{figure=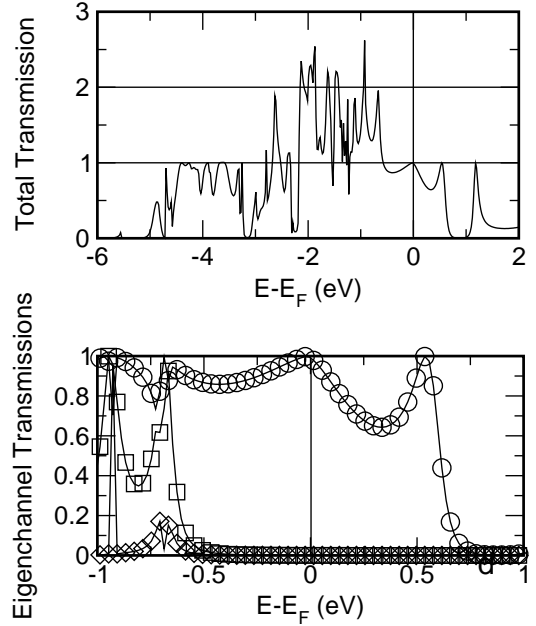, width=0.8\columnwidth, angle=0}
\caption{The total transmission of all channels 
and eigenchannel transmissions of the 5 atom long chain 
shown in Fig.~\ref{fig:5struc}.}
\end{figure}

\subsubsection{Finite bias results}

Most experimental studies of atomic wires have been done in the low
voltage regime ($V < 0.25$ V). Important questions about the nonlinear
conductance, stability against electromigration and heating effects
arises in the high voltage regime. It has been found that the single
atom gold wires can sustain very large current densities, with an
intensity of up to 80$\mu$A corresponding to 1 V bias.\cite{YaBoBr98}
Sakai and co-workers\cite{YaSa97,ItYuKu99,YuEnSa01} have measured the
conductance distributions (histograms) of commercial gold relays at
room temperature and at 4K and found that the prominent 1 $G_0$ peak
height decreases for high biases $V > 1.5$ V and disappears around 2
V. It is also observed that there is no shift in the 1 $G_0$ peak
position which indicates that the nonlinear conductance is small. In
agreement with this Hansen {\it et al.}\cite{HaNiBr00} reported linear
current-voltage (\IV) curves in STM-UHV experiments and suggested that
nonlinearities are related with presence of contaminants. On the
theoretical side $s,p,d$-tight-binding
calculations\cite{BrKoTs99,HaNiBr00} has been performed for voltages
up to 2.0 V for atomic gold contacts between (100),(111) and (110)
electrodes.  Todorov {\it et al.}\cite{ToHoSu00,ToHoSu01} has
addressed the forces and stability of single atom gold wires within a
single orbital model combined with the fixed atomic charge condition.

\begin{figure}
  \epsfig{figure=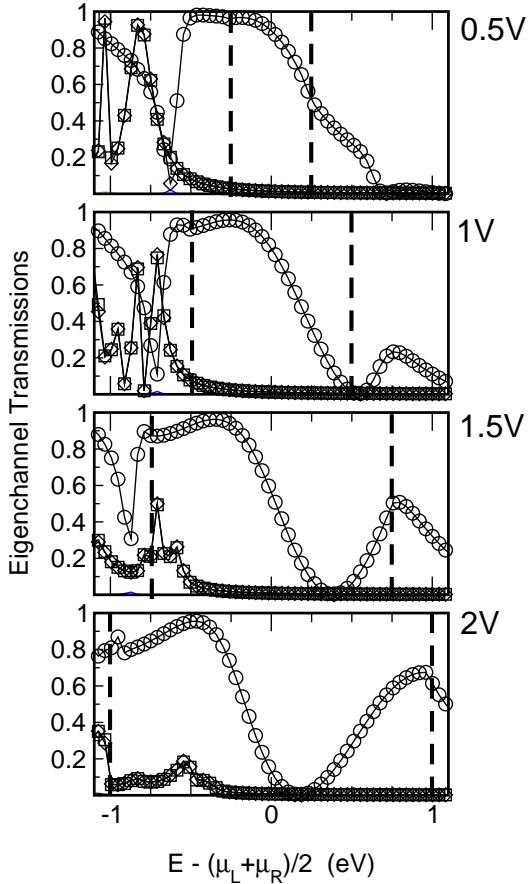, width=0.8\columnwidth, angle=0}
\caption{The eigenchannel transmissions for bias voltages of 
  0.5 V, 1 V, 1.5 V, and 2 V for the wire shown in Fig.~\ref{fig:3struc}.
  The conductance is determined from the average total transmission
  from $\mu_L$ to $\mu_R$.  The voltage window is shown with thick
  dashed lines.}
\label{fig:teigsvolt}
\end{figure}

Here we study the influence of such high currents and fields on one of
the wire structures ((c) in Fig.~\ref{fig:3struc}).  We have performed
the calculations for voltages from 0.25 V to 2.0 V in steps of 0.25 V. In
Fig.~\ref{fig:teigsvolt} we show the eigenchannel transmissions for
finite applied bias. For a bias of 0.5 V we see a behavior similar to
the 0 V situation except for the disappearance of the resonance
structure about 0.7 eV above $E_F$ in Fig.~\ref{fig:allteig}(c).  For
0.5 V bias the degenerate peak 0.75 eV below $E_F$ which is derived from
the $l_z=1$ orbitals is still intact whereas this feature diminishes
gradually for higher bias. Thus mainly a single channel contributes
for finite bias up to 2 V. It is clear from Fig.~\ref{fig:teigsvolt}
that the transmissions for zero volts cannot be used to calculate the
conductance in the high voltage regime and underlines the need for a
full self-consistent calculation.

\begin{figure}
  \epsfig{figure=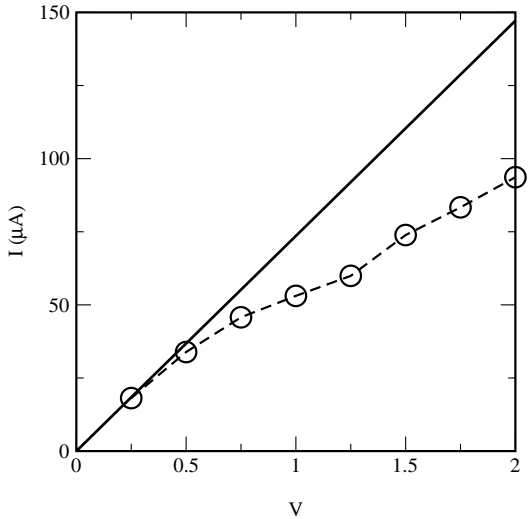, width=0.8\columnwidth, angle=0}
\caption{The current-voltage (\IV) curve for the wire shown in Fig.~\ref{fig:3struc}.}
\label{fig:iv}
\end{figure}

The calculated \IV curve is shown in Fig.~\ref{fig:iv}. We observe a
significant decrease in the conductance ($I/V$) for high voltages.
This is in agreement with tight-binding results\cite{BrKoTs99} where a
30\% decrease was observed for a bias of 2 V. For wires
attached to (100) and (110) electrodes\cite{BrKoTs99,HaNiBr00} a quite
linear \IV was reported for the same voltage range. 

\begin{figure}
  \epsfig{figure=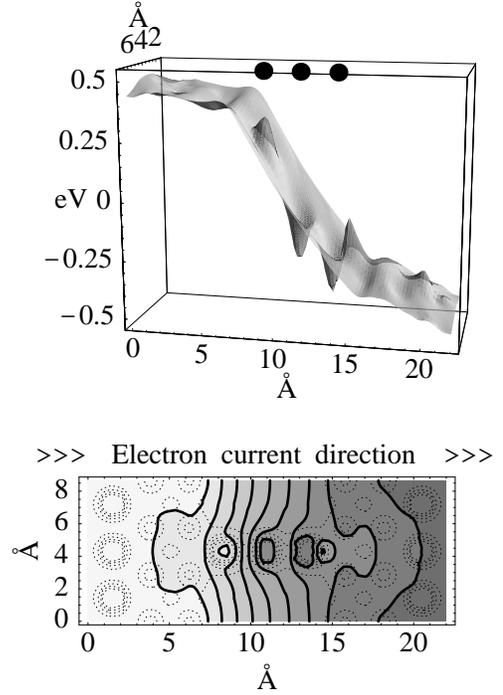, width=0.8\columnwidth, angle=0}
\caption{The voltage drop for applied bias of 1 V in a plane going through
  the wire atoms. In the surface plot the wire atom positions are
  shown as black spheres. In the contour plot below the solid contours
  (separated by 0.1 eV) show the voltage drop.  The dashed contours are
  shown to indicate the atomic positions.}
\label{fig:vdrop1V}
\end{figure}

\begin{figure}
  \epsfig{figure=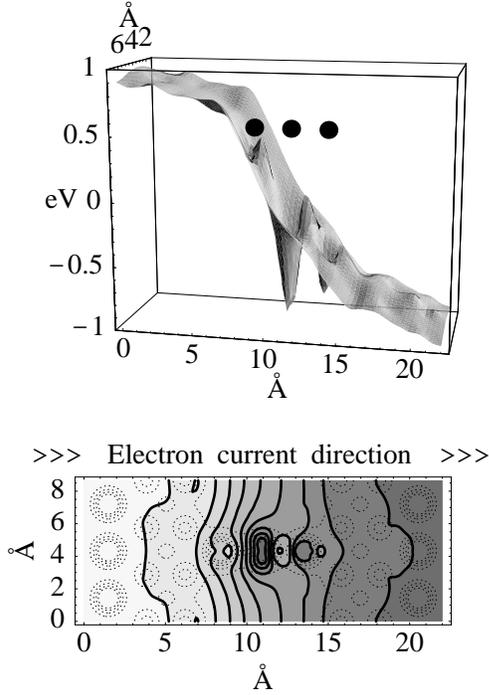, width=0.8\columnwidth, angle=0}
\caption{Same as in Fig.~\ref{fig:vdrop1V} for a bias of 2 V (contours separated by 0.2 eV).}
\label{fig:vdrop2V}
\end{figure}

In Fig.~\ref{fig:vdrop1V} and Fig.~\ref{fig:vdrop2V} we plot the
voltage drop - {\it i.e.}  the change in total potential between the
cases of zero and finite bias, for the case of 1 V and 2 V,
respectively.  We observe that the potential drop has a tendency to
concentrate in between the two first atoms in the wire in the direction
of the current. A qualitatively similar behavior was seen in the
tight-binding results for both (100) and (111)
electrodes\cite{BrKoTs99} and it was suggested to be due to the
details of the electronic structure with a high density of states just
below the Fermi energy derived from the $d$ orbitals (and their
hybridization with $s$ orbitals). The arguments were based on the
atomic charge neutrality assumption. In the present calculations,
this assumption is not made, although the selfconsistency and
the screening in the metallic wire will drive the electronic
distribution close to charge neutrality. This would not occur in
the case of a non-metallic contact.\cite{PaPeLoVe01,TaGuWa01a}

\begin{figure}
  \epsfig{figure=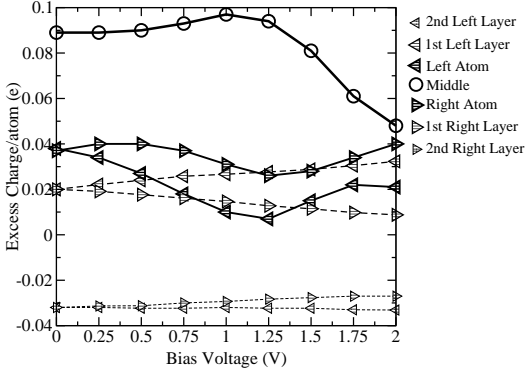, width=0.8\columnwidth, angle=0}
\caption{The (Mulliken) change in excess charge (in units of the electron charge) 
  on the wire atoms and average excess charge on the surface atoms in
  the first and second left and right electrode layers.  For high bias
  the 2nd right electrode layer takes up some of the excess charge.  }
\label{fig:chargelayer}
\end{figure}

The number of valence electrons on the gold atoms is close to 11.
There is some excess charge on the wire atoms and first surface layers
(mainly taken from the second surface layers).  The behavior of the
charge with voltage is shown in Fig.~\ref{fig:chargelayer}.  The
minimum in voltage drop around the middle atom for high bias (see
Fig.~\ref{fig:vdrop2V}) is associated with a decrease in its excess
charge for high bias. The decrease is found mainly in the $s$ and
$d_{zz}$ orbitals of the middle atom.

\subsubsection{Forces for finite bias}

\begin{figure}
  \epsfig{figure=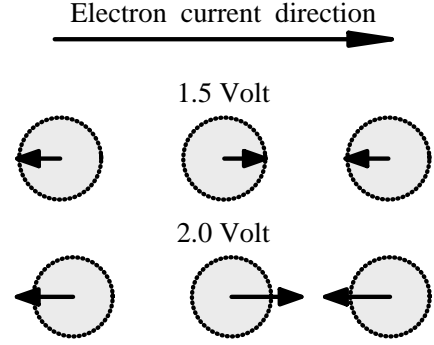, width=0.7\columnwidth, angle=0}
\caption{The forces acting on the wire atoms when the bias is applied 
(the radius of the circles correspond to 0.5 nN). The tensile force in the bond 
between the two first atoms is about 1 nN for 1 V and 1.5 nN for 2 V.}
\label{fig:forces}
\end{figure}

We end this section by showing the forces acting on the three atoms in
the wire for finite bias in Fig.~\ref{fig:forces}.  We evaluate the
forces for non-equilibrium in the same manner as for equilibrium
{\sc Siesta} calculations\cite{OrArSo96} by just using the nonequilibrium
density matrix and Hamiltonian matrix instead of the equilibrium
quantities.\cite{ToHoSu00} 
We find that for voltages above 1.5 V that the first bond
in the chain wants to be elongated while the second bond wants to
compress. Thus the first bond correspond to a ``weak spot'' as
discussed by Todorov {\it et al.}\cite{ToHoSu00,ToHoSu01} We note that
the size of the bias induced forces acting between the two first wire
atoms at 2 V is close to the force required to break single atom
contacts\cite{RuBaAg01} (1.5$\pm$0.3 nN) and the result therefore
suggests that the contact cannot sustain a voltage of this magnitude,
in agreement with the relay experiments.\cite{YuEnSa01} A more
detailed calculation including the relaxation of the atomic
coordinates for finite voltage bias is needed in order to draw more firm
conclusions about the role played by the nonequilibrium forces on the
mechanical stability of the atomic gold contacts. 
We will not go further into the analysis of the electronic structure
and forces for finite bias in the gold wire systems at this point,
since our aim here is simply to present the method and show some of its 
capabilities. A full report of our calculations will be published elsewhere.

\subsection{Conductance in nanotubes}
Finally, we have applied our approach to the calculation of
conductance of nanotubes in the presence of point defects. In
particular the Stone-Wales (SW) defect\cite{stonewalles} ({\em i.e.},
a pentagon-heptagon double pair) and a monovacancy in a $(10,10)$
nanotube. The atomic geometries of these structures are obtained from
a {\sc Siesta} calculation with a 280-atom supercell (7 bulk unit
cells), where the ionic degrees of freedom are relaxed until any
component of the forces is smaller than 0.02 eV/\AA.  We use a
single-$\zeta$ basis set, although some tests were made with a
double-$\zeta$ basis, producing very similar results.  The one
dimensional Brillouin zone is sampled with five k-points. The forces
do not present any significant variation if the the relaxed
configurations are embedded into a 440-atom cell, where the actual
transport calculations are performed.

\begin{figure}
\centerline{\epsfig{file=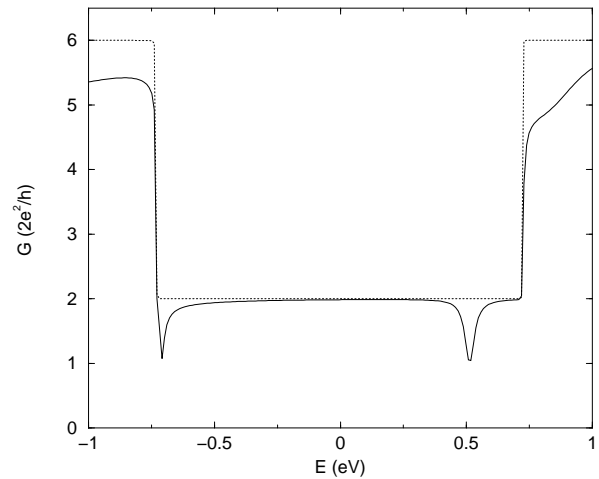,width=0.9\columnwidth}}
\caption[]{
  Transmission coefficient of pentagon-heptagon double pair vs. energy
  measured with respect to the Fermi level. The dotted line shows the
  transmission of a perfect nanotube.  }
\label{5775}
\end{figure}

In a perfect nanotube two channels, of character $\pi$ and $\pi^*$,
contribute each with a quantum of conductance, $G_0$.  In Fig.~\ref{5775} 
we present our results for zero bias for the SW defect.
Recent {\em ab initio} studies \cite{ChIh99,ChIhLoCo00} are well
reproduced, with two well defined reflections induced by defect
states. The two dips in the conductance correspond to the closure of
either the $\pi^*$ (below the Fermi level) or the $\pi$ channel.

\begin{figure}
  (a) \centerline{\epsfig{file=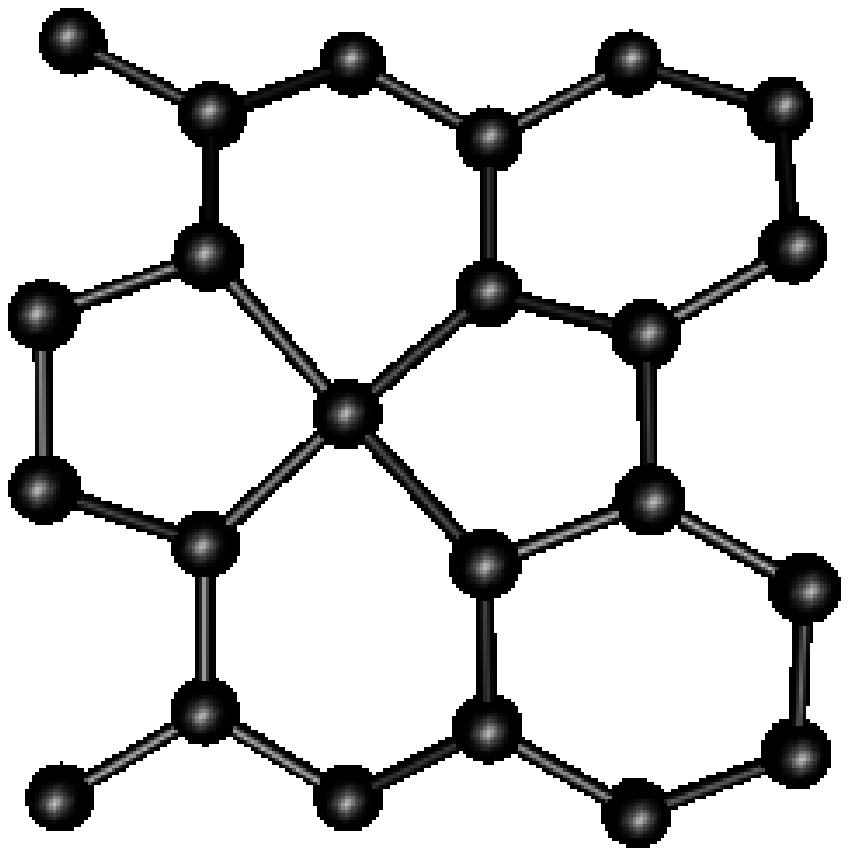,width=0.5\columnwidth}}
  
  (b) \centerline{\epsfig{file=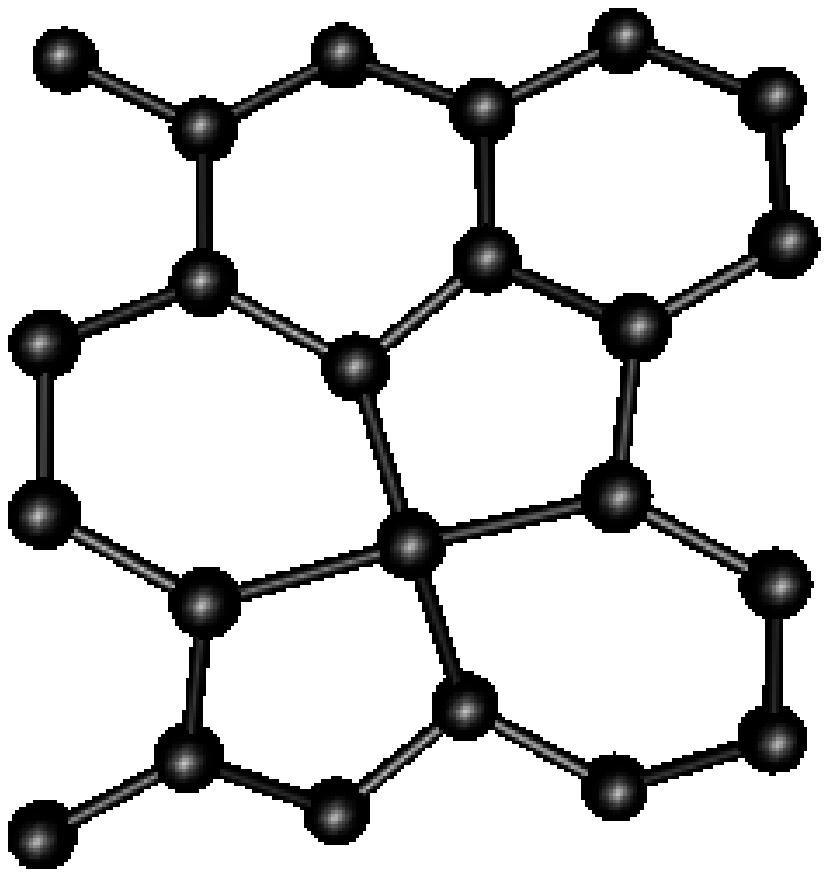,width=0.5\columnwidth}}
\caption[]{
  Atomic configurations for the vacancy defect in a (10,10) nanotube:
  (a) metastable and (b) ground state.  }
\label{figvac}
\end{figure}

For the ideal vacancy the two antibonding states associated with
broken $\sigma$ bonds lie close to the Fermi level.  The coupling
between these states and the $\pi$ bands, although small, suffices to
open a small gap in the bulklike $\pi-\pi^*$ bands.  The
vacancy-induced states appear within this gap.  Otherwise, the three
two-coordinated atoms have a large penalty in energy and undergo a
large reconstruction towards a split vacancy configuration with two
pentagons, $\sim 2$~eV lower in energy.  Two configurations are
possible, depending on the orientation of the pentagon pair, depicted
in Fig.~\ref{figvac}. We have found that there is a further $0.4$~eV
gain in energy by reorienting the pentagon-pentagon $60^o$ off the
tube axis (Fig.~\ref{figvac}b) resulting in a formation energy of $E_f
= 6.75$~eV.\cite{ef-def} The bonding of the tetracoordinated atom is
not planar but paired with angles of $\sim 158^0$.  Some of these
structures were discussed in previous tight-binding
calculations.\cite{AjRaCh98} This is at variance with the results of
Refs.~\onlinecite{ChIh99,ChIhLoCo00}, possibly due to their use of too
small a supercell which does not accommodate the long range elastic
relaxations induced by these defects.

\begin{figure}
  \centerline{\epsfig{file=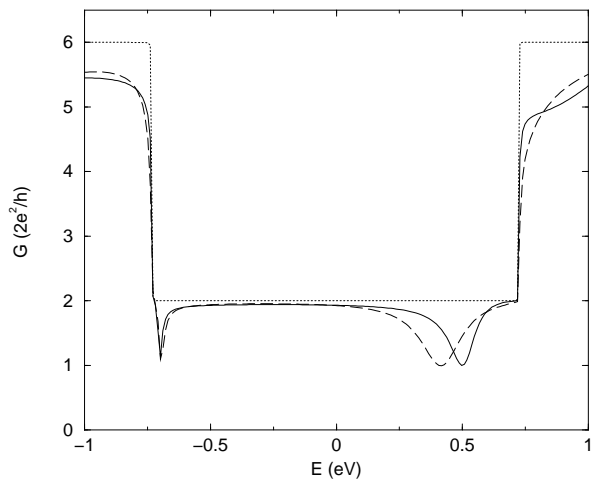,width=0.9\columnwidth}}
\caption[]{
  Similar to Fig. \ref{5775} but now for the vacancy: ground
  (continuous line) and metastable (dashed line) configurations.  }
\label{condvac}
\end{figure}

The conductance of these defects, calculated at zero bias
(Fig.~\ref{condvac}), does not present any features close to the Fermi
level. This is in contrast to the ideal vacancy, where reflection
related to the states mentioned above are present.  Two dips appear,
at possitions similar to those of the SW deffect. An eigenchannel
analysis\cite{BrSoJa97} of the transmission coefficients gives the
symmetry of the states corresponding to these dips.  The metastable
configuration is close to having a mirror plane, containing the tube
axis, except for the small pairing mode mentioned before.  The mixing
of the $\pi$ and $\pi^*$ bands is rather small.  The lower and upper
dips come from the reflection of the almost pure $\pi^*$ and $\pi$
eigenchannels, respectively.  This behavior is qualitatively similar
to the SW defect. On the other hand, for the rotated pentagon pair
there is no mirror plane and the reflected wave does not have a well
defined character.

\section{Conclusion}
We have described a method and its implementation({\sc TranSiesta}) for
calculating the electronic structure, electronic transport, and forces
acting on the atoms at finite voltage bias in atomic scale systems.
The method deals with the finite voltage in a fully self-consistent
manner, and treats both the semi-infinite electrodes and the contact
region with the same atomic detail.

We have considered carbon wires connected to aluminum electrodes where
we find good agreement with results published earlier with another
method (McDCAL)\cite{LaTaMeGu01} for electrodes with finite cross
section. We find that the voltage drop through the wire system depends
on the detailed structure of the electrodes ({\it i.e.}  periodic
boundary conditions vs. cross section).

The conductance of three and five atom long gold wires with a bend
angle has been calculated. We find that the conductance is close to
one quantum unit of conductance and that this result is quite stable
against the bending of the wire. These results are in good agreement
with experimental findings. For finite bias we find a non-linear
conductance in agreement with previous semi-empirical calculations for
(111) electrodes.\cite{BrKoTs99} We find that the forces at finite
bias are close to the experimental force needed to break the gold
wires\cite{RuBaAg01} for a bias of 1.5 to 2.0 V.

Finally we have studied the transport through a (10,10) nanotube with
a the Stone-Wales defect or with a monovacancy (a calculation
involving 440 atoms). We have found good agreement with recent {\it ab
  initio} studies of these systems.\cite{ChIh99,ChIhLoCo00}

\begin{acknowledgments} 
We thank Prof.\ Hans Skriver for sharing his insight
into the complex contour method with us, Prof. J. M. Soler for useful
comments on the electrostatic potential problem, and Prof. A.-P. Jauho
for discussions on the non-equilibrium method. This work has
benefited from the collaboration within, and was partially funded by,
the ESF Programme on ``Electronic Structure Calculations for
Elucidating the Complex Atomistic Behaviour of Solids and Surfaces".
We acknowledge support from the Danish Research Councils (M.B. and
K.S.), and the Natural Sciences and Engineering Research Council
(NSERC) (J.T.).  M.B. has benefited from the European Community - 
Access to Research Infrastructure action of the Improving Human
Potential Programme for a research visit to the ICMAB and CEPBA
(Centro Europeo de Paralelismo de Barcelona).  J.L.M.  and P.O.
acknowledge support from the European Union (SATURN IST-1999-10593),
the Generalitat de Catalunya (1999 SGR 207), Spain's DGI
(BFM2000-1312-C02) and Spain's Fundaci\'on Ram\'on Areces.

Part of the calculations were done using the computational
facilities of CESCA and CEPBA, coordinated by C$^4$.
\end{acknowledgments}

\bibliographystyle{prsty}

\begin{thebibliography}{10}
\bibitem{notepoisson}
Our computational cell is constructed to include to a very good approximation
  all the screening effects on the metallic leads. Therefore, beyond that, the
  charge distribution is that of a bulk lead. Then a linear ramp is the
  appropiate homogeneous solution of the Poisson equation to be considered
  here.
\bibitem{notepoisson2}
This potential is defined up to an arbitrary constant, which is fixed in
  practice by fixind the average value at the $z=0$ cell boundary.
\bibitem{trans-mcdcal}
In Tran{\sc Siesta}, the Hamiltonian parameters are fixed to their
  corresponding bulk values, while, in McDCAL, the Hamiltonian parameters that
  describe the electrodes are calculated by fixing the real space effective
  potential to the corresponding bulk value. In Tran{\sc Siesta}, the Poisson
  equation is solved using a FFT technique with periodic boundary conditions,
  while, in McDCAL, it is solved in real space by requiring that the Hartree
  potential be equal to the bulk values of the left and right boundaries of the
  simulation cell. In Tran{\sc Siesta}, the electron charge is obtained using
  the double contour technique described in Sec.~\ref{sec:nedm}, while a single
  contour technique is used in McDCAL.
\bibitem{siestaprecalc}
We note here that, of course, our formalism does not deliver the density matrix
  values for the auxiliary buffer region. These are, however, required to
  construct the density at the scattering region. We obtained them from a
  initial periodic {\sc Siesta} calculation on the same cell and we fixed them
  hereon. A subsequent zero bias {\sc TranSiesta} calculation yields the same
  results within numerical accuracy.
\bibitem{ef-def}
The formation energy was calculated as $E_f = E_d (N) - N \mu_C$. $E_d $ is the
  total energy of the supercell of $N$ atoms containing the defect. The carbon
  chemical potential, $\mu_C$, corresponds to the perfect (10,10) nanotube.
\end{thebibliography}

\end{document}